\documentclass{article}

\usepackage{authblk,amsmath,amssymb,graphicx,geometry,cite,color,mathptmx,courier,textcomp,verbatim}
\usepackage[table]{xcolor}

\begin{document}

\title{Network of Time-Multiplexed Optical Parametric Oscillators as a Coherent Ising Machine}

\author{Alireza Marandi$^{1,2}$, Zhe Wang$^{1}$, Kenta Takata$^{2,3}$, \\ Robert L. Byer$^{1}$, and Yoshihisa Yamamoto$^{1,2,3}$}
\affil{$^1$ E. L. Ginzton Laboratory, Stanford University, California 94305, USA,}
\affil{$^2$ National Institute of Informatics, Tokyo 101-8403, Japan,}
\affil{$^3$ Department of Information and Communication Engineering, University of Tokyo, Tokyo 113-8656, Japan.}
\maketitle

{\bf Finding the ground states of the Ising Hamiltonian \cite{ising_problem} maps to various combinatorial optimization problems in biology, medicine, wireless communications, artificial intelligence, and social network. So far no efficient classical and quantum algorithm is known for these problems, and intensive research is focused on creating physical systems - Ising machines - capable of finding the absolute or approximate ground states of the Ising Hamiltonian \cite{simulated_annealing, quantum_annealing, quantum_annealing2, QAC, dwave}. Here we report a novel Ising machine using a network of degenerate optical parametric oscillators (OPOs). Spins are represented with above-threshold binary phases of the OPOs and the Ising couplings are realized by mutual injections \cite{zhe}. The network is implemented in a single OPO ring cavity with multiple trains of femtosecond pulses and configurable mutual couplings, and operates at room temperature. We programed the smallest non-deterministic polynomial time (NP)- hard Ising problem on the machine, and in 1000 runs of the machine no computational error was detected. }

Many important combinatorial optimization problems belong to the NP-hard or NP-complete classes, and solving them efficiently has been beyond the reach of classical and quantum computers. Currently, simulated annealing \cite{simulated_annealing} and various approximate algorithms \cite{comb, approx, approx2} are among the widely used methods in classical digital computers. Recently, quantum annealing \cite{quantum_annealing, quantum_annealing2, dwave} and adiabatic quantum computation \cite{QAC} have been studied intensively for solving those NP problems. Even though relatively large machines are recently implemented to perform quantum annealing \cite{dwave}, their comprehensive potentials are yet to be explored \cite{qac_th, dwave_2013}, and the quest for a novel computing machine to solve large-scale NP-hard problems continues.

Recently, degenerate OPOs pumped by femtosecond laser pulses have been shown useful for a wide range of classical and quantum applications. While operated above their oscillation threshold, they are used in generation of broadband mid-infrared frequency combs which are intrinsically phase and frequency locked to the pump \cite{fc_opo}. Below threshold, they are employed for realization of quantum networks using wavelength division multiplexing \cite{wdm_network}. Intriguingly, these OPOs demonstrate non-equilibrium phase transition at threshold, which has been previously exploited for optical quantum random number generation \cite{rngpaper}. Here we explore the computational capability of this phase transition in a coherent network of time-division-multiplexed femtosecond OPOs.

The Hamiltonian of an Ising model with $N$ spins and without an external field is given by \cite{ising}:
\begin{equation}
{H}=-\sum_{ij}^{N}J_{ij} \sigma_i \sigma_j,
\end{equation}
with $J_{ij}$ being the coupling between the $i^{th}$ and $j^{th}$ spins, and $\sigma_i$ and $\sigma_j$ representing the $z$-projections of the spins, the eigenvalues of which are either +1 or -1. To implement an artificial Ising spin system, elements with binary degrees of freedom and configurable couplings are required. This has been demonstrated experimentally in cold ion or atom system \cite{trapped_ion, optical_lattice } with practical difficulties in scalability. Artificial Ising spin systems based on a network of injection locked lasers are also proposed \cite{kenta, shoko}. 

\begin{figure}[tb]
\centering
\includegraphics[width=.9\textwidth]{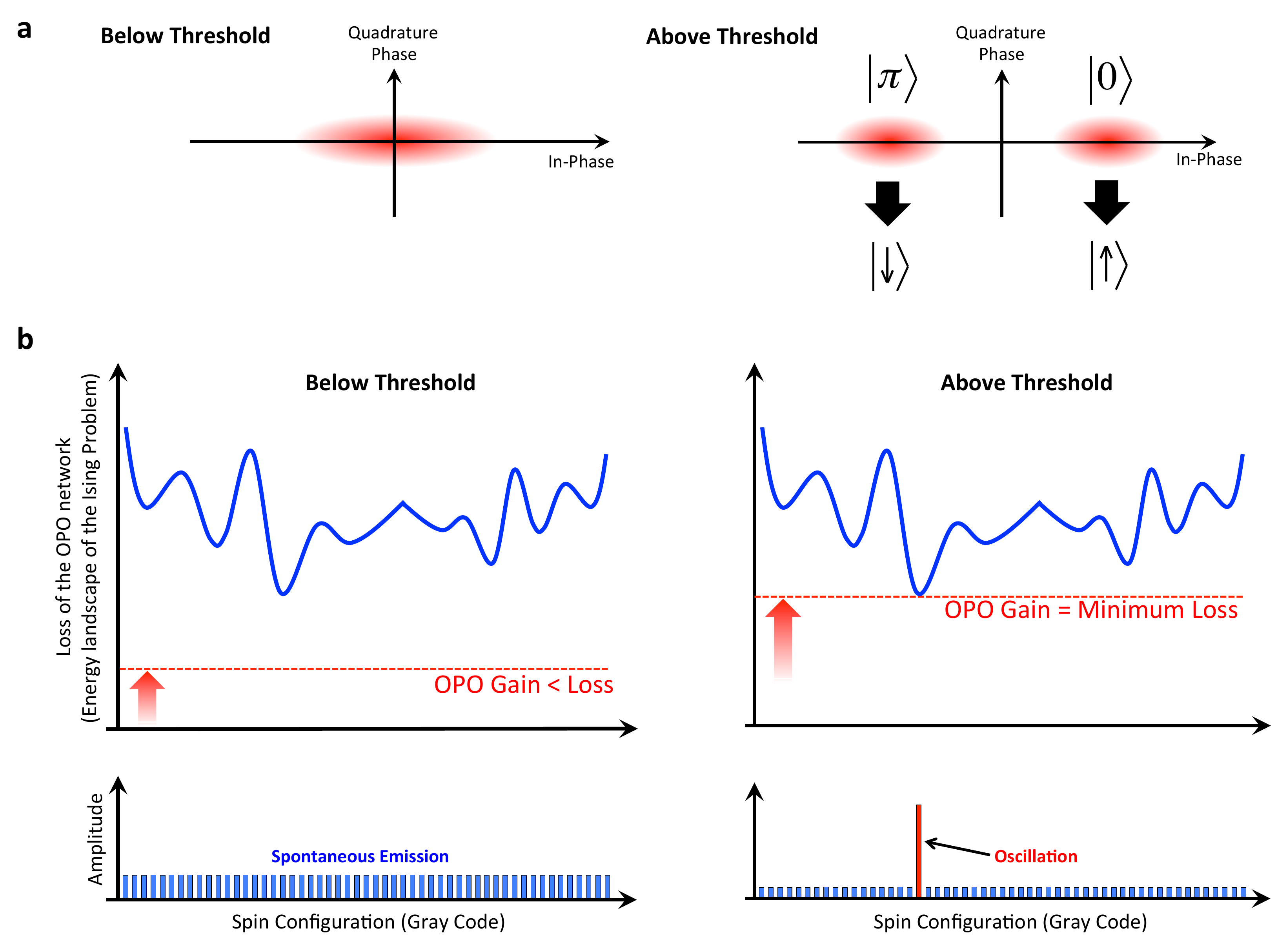}
\caption{{\bf Principle of operation of OPO Ising machine.} {\bf a,} Simplified illustration of vacuum squeezing in a degenerate OPO below threshold, and the binary phase states above threshold shown in the in-phase and quadrature-phase coordinates. {\bf b,} Gradual pumping of the OPO network to find the ground states. The parametric gain is gradually increased from below threshold to reach the minimum loss of the ground state. Since the lowest loss of the network corresponds to the phase-state configuration representing the answer of the Ising problem, it is expected that only the ground state oscillates. }
\label{fig:operation_principle}
\end{figure}

Degenerate OPOs are particularly suitable to represent Ising spins because of binary phase operation above threshold \cite{nabors}. Fig. \ref{fig:operation_principle}a illustrates the operation of a degenerate OPO for below and above oscillation threshold in the in-phase and quadrature-phase coordinates. Below the threshold, the signal field is a squeezed vacuum state \cite{squeezing}, and by increasing the pump field it undergoes spontaneous symmetry breaking around the oscillation threshold \cite{symmetry_breaking}. This non-equilibrium phase transition leads to oscillation in one of the two phase states ($|0\rangle$ or $|\pi\rangle$) \cite{nabors}, which are exploited to represent an Ising spin ($\sigma_i$). The couplings between the spins ($J_{ij}$) are realized by mutual injections of the signal fields of the $i^{th}$ and $j^{th}$ OPOs. The resulting OPO network has a phase-state dependent photon decay rate corresponding to the energy landscape of the original Ising Hamiltonian \cite{zhe}, i.e. different spin configurations (phase states) have different total photon losses in the network, which is illustrated in Fig. \ref{fig:operation_principle}b. 

The search for the ground states is conducted by gradually raising the gain via increasing the pump field, as illustrated in Fig. \ref{fig:operation_principle}b. As the parametric gain approaches the lowest possible loss, the network goes through the OPO phase transition, and as a result it is expected to oscillate in one of the exact or approximate ground states. Further increase in the pump pins the gain to the threshold value and thus erroneous oscillation in the excited states continues to be suppressed. Although the below or above-threshold behavior of an OPO network can be efficiently simulated on a classical computer, no efficient algorithm is yet known for comprehensive simulation of the phase transition point. 

We use a numerical model based on c-number Langevin equations, which are compatible with fundamental quantum mechanical master equations or Fokker-Planck equations. We simulate OPO networks of up to 20000 OPOs to solve some of the NP-hard maximum cut (MAX-CUT) problems. For these problems, the OPO network outperforms one of the well-known approximate algorithms in terms of the computation time and the accuracy of the results (see Supplementary Information). However, despite these promising numerical results, no mathematical proof exists to guarantee superior performance of the OPO network for all Ising problems, and the possibility of failure for some instances exists.

\begin{figure}[htbp]
\centering
\includegraphics[width=.7\textwidth]{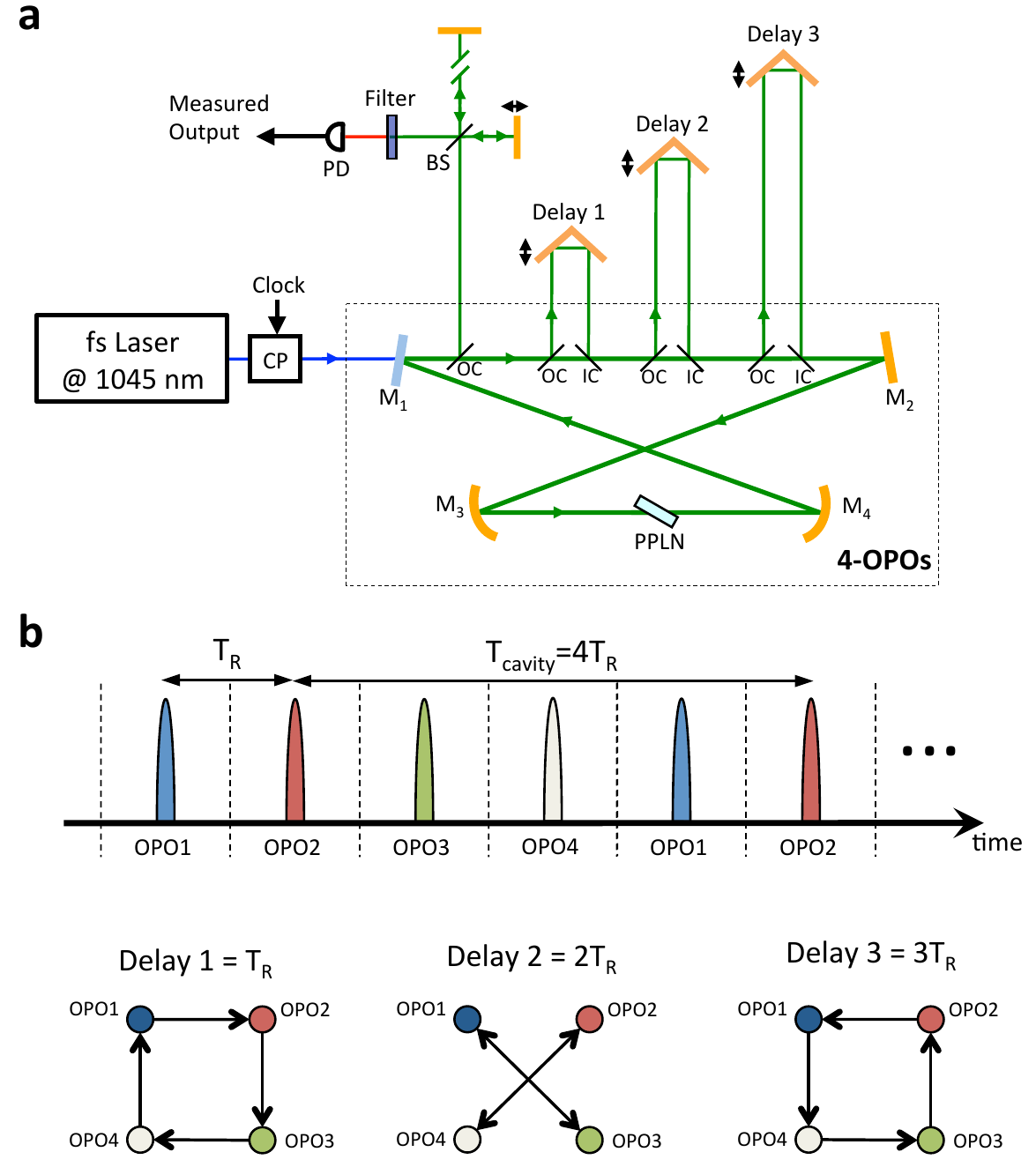}
\caption{ {\bf Experimental setup.}   {\bf a}, The setup composed of a 4-OPO system pumped by a femtosecond fiber laser at 1045 nm, three delay lines to provide couplings between the OPOs, and an unequal-arm Michelson interferometer for measurement of the relative phase states of adjacent OPOs. A chopper (CP) restarts the OPO system periodically with a clock rate of $\sim$1 kHz and a rise time (10-90\% power) of $\sim$180 $\mu$s. Output couplers (OCs) and input couplers (ICs) provide $\sim$4\% of power reflection (varying between 2 and 6\%). The nonlinear crystal is periodically poled lithium niobate (PPLN).  {\bf b}, Illustration of the output pulse train and the time slots assigned to OPOs 1 to 4. The repetition period of the pump and the OPO output ($T_R$) is 4 ns, and the cavity roundtrip time ($T_{cavity}$) is 16 ns. Each delay line provides four couplings, i.e. two-body interactions, among the temporally separated OPOs. Delay 1 couples OPO$_n$ to OPO$_{n+1}$ ($J_{12}$, $J_{23}$, $J_{34}$, $J_{41}$), delay 2 couples OPO$_n$ to OPO$_{n+2}$ ($J_{13}$, $J_{24}$, $J_{31}$, $J_{42}$), and delay 3 couples OPO$_n$ to OPO$_{n+3}$ ($J_{14}$, $J_{21}$, $J_{32}$, $J_{43}$). }
\label{fig:sch}
\end{figure} 

For the experiment, we exploit the novel scheme of ``time-division-multiplexing", in which a single ring resonator accommodates many Ising spins simultaneously, which are free from mismatch and phase decoherence noise. Figure \ref{fig:sch}a shows the experimental setup comprising a 4-OPO system. The ring resonator has a roundtrip time ($T_{cavity}$) of four times the pump pulse repetition period ($T_R$), as shown in Fig. \ref{fig:sch}b, and therefore the pulses represent four temporally separated independent OPOs. The couplings between these independent but identical OPOs are realized using three pairs of output and input couplers in the resonator paths. In each pair, $\sim$4\% of the intracavity power is out-coupled and precisely delayed in a free-space delay line; each delay is an integer multiple of the repetition period, and $\sim$4\% of the out-coupled light is injected back to the resonator. This corresponds to $\sim$4\% of field coupling between the OPOs. Figure \ref{fig:sch}b depicts how these delay lines provide all the possible couplings ($J_{ij}$) among the four temporally separated OPOs.

To measure the phase states of the 4-OPO system, the output is sent to a one-bit delay Michelson interferometer as shown in Fig. \ref{fig:sch}a. The delay difference of the arms is locked to the repetition period of the system ($T_R$). Therefore, the interferometer measures the differential phase between adjacent OPOs, i.e. the output is the interference of OPO$_n$ and OPO$_{n+1}$, and $n$ changes with the time slot of measurement. We use fast and slow detectors at the output of the interferometer and in Supplementary Information we show the expected outputs for different phase states.

The 4-OPO system is pumped with a femtosecond mode locked laser with a central wavelength of 1045 nm,  and a repetition period of 4 ns. It operates at degeneracy and the detailed characterization is presented in Supplementary Information. The pump field is gradually increased from below to about two times above the oscillation threshold over a time of $\sim$180 $\mu$s, while the cavity photon lifetime is $\sim$60 ns. 

\begin{figure}[htbp]
\centering
\includegraphics[width=\textwidth]{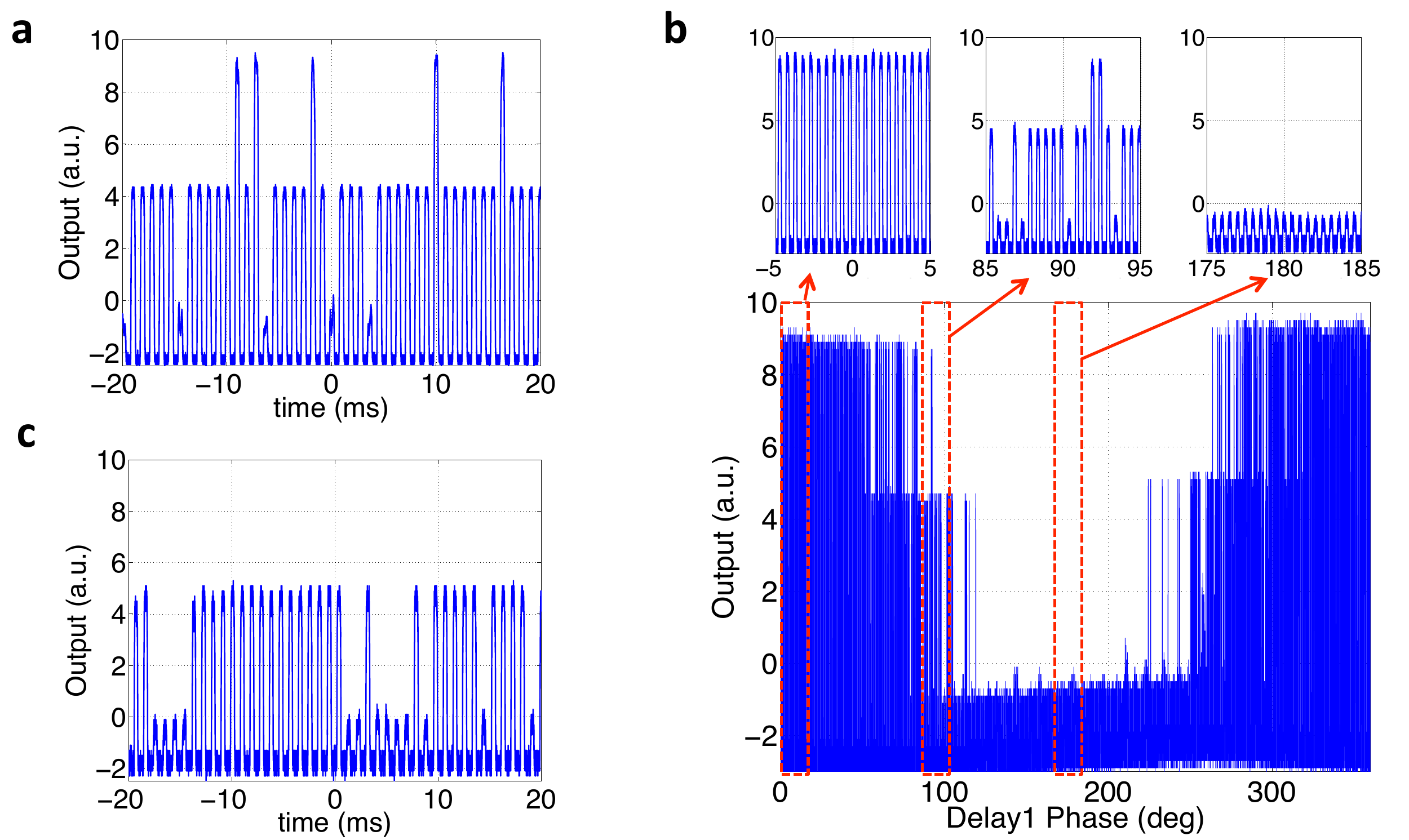}
\caption{{\bf Slow detector results}. {\bf a}, All delay lines are blocked. The output is randomly at one of the three discrete intensity levels of $0$, $I_m/2$, and $I_m$. The reason that the $0$ intensity level is slightly above the noise floor is mainly because the arms of the interferometer have different lengths and therefore the interfering beams are not completely overlapping.  {\bf b}, Only delay 1 is present and its phase is scanned slowly. Around zero, where the coupling is in phase, the output locks to $I_m$ verifying that all OPOs are in the same phase states. Around 180 degrees, where the coupling is out of phase, the output locks at $0$ verifying that adjacent OPOs have opposite phase states. The three-state output around 90 degrees shows the inefficiency of coupling at those phases. Regenerative behavior due to the phase sensitive amplification of the OPO is observed for example for the coupling phases between -30 and 30 degrees and between 150 and 210 degrees. {\bf c}, All the delays are present and their phases are locked to 180 degrees to represent MAX-CUT problem, where the output toggles between $0$ and $I_m/2$. }
\label{fig:slow_det}
\end{figure}

First, we present the measurements using a slow detector in Fig. \ref{fig:slow_det}. When the couplings are blocked, the OPOs are expected to operate independently, and therefore a random uniformly distributed selection of the phase states are expected each time the OPO is restarted. A sample of the interferometer output for this case is shown in Fig. \ref{fig:slow_det}a. The output toggles between the three distinct intensity levels (0, $I_m/2$, and $I_m$).  Assuming all the states are equally probable to occur, $I_m/2$ corresponds to 12 phase states and $I_m$ and $0$ correspond to 2 different phase states (Supplementary Information). Indeed the frequency of these events are 6:1:1, as shown in Fig. \ref{fig:slow_det}a.  

Having in-phase (out-of-phase) couplings among the adjacent OPOs corresponds to ferromagnetic (antiferromagnetic) couplings in an Ising spin ring with the expected outcome of aligned (anti-aligned) spins. In the experiment, when the shortest delay line is set to in-phase (out-of-phase) coupling, the OPOs operate at the same (alternating) phase states. We show this behavior in Fig. \ref{fig:slow_det}b where the phase of delay 1 is swept and the other delays are blocked. The sweeping speed is slow compared with the 1 kHz restarting frequency of the OPOs. The output of the slow detector is always at $I_m$ when phase of delay 1 is around zero ($ J_{12}= J_{23}= J_{34}= J_{41}> 0$). This corresponds to $|0000\rangle$ or $|\pi\pi\pi\pi\rangle$ phase states confirming that in-phase coupling between OPO$_n$ and OPO$_{n+1}$ leads to all OPOs oscillating in the same phase state. When the phase of delay 1 is around 180$^\circ$, i.e. the field of OPO$_n$ is shifted by $\pi$ and injected to OPO$_{n+1}$, the output intensity remains at $0$ level, meaning that the OPOs are either in $|0\pi0\pi\rangle$ or $|\pi0\pi0\rangle$. As indicated in the plot, the coupled OPOs tolerate a coupling phase deviation of at least $\pm$30$^\circ$ around these two points of in-phase or out-of-phase couplings. This regenerative behavior is due to the phase sensitive nature of the parametric gain in the degenerate OPO and makes the system immune to the phase noise in the environment. Similar behavior is observed for the other delay lines and the detailed results are presented in Supplementary Information.

We focus on implementation of an NP-hard MAX-CUT problem, which is mapped to an Ising problem \cite{maxcut_ising} and corresponds to a frustrated Ising spin system \cite{trapped_ion}. The problem is to find a subset of vertices in a cubic graph, in which all vertices have three edges, such that the number of edges between the subset and its complementary subset is maximized. For a cubic graph with 4 vertices, which has 6 edges, any subset choice with two vertices is an answer of the MAX-CUT problem (or MAX-2-SAT problem), and any other subset choice is not. To program this problem on the OPO Ising machine, all the mutual couplings between the OPOs are set to out of phase, i.e. $ J_{ij}<0 $. The total photon decay rate of the network is lowest for the answers of the MAX-CUT problem \cite{zhe} corresponding to two OPOs in $|0\rangle$ and two OPOs in $|\pi\rangle$, hence upon each gradual pumping the 4-OPO network is expected to oscillate in such a phase state.

The phases of the three delays are locked to $\pi$ using the residual pump interference signal at the transmission port of the input couplers, and a sample of the interferometer output is shown in Fig. \ref{fig:slow_det}c. It matches the expected outcome of being either $I_m/2$ or $0$ with the rate of $I_m/2$ twice the rate of $0$. However, since $I_m/2$ corresponds to both answer and non-answer states, it is necessary to use a fast detector to confirm that the non-answer states do not occur. Other configurations for the coupling phases (corresponding to other Ising problems) are also investigated using the slow detector and the results follow the expected outcomes (see Supplementary Information).

\begin{figure}[htbp]
\centering
\includegraphics[width=\textwidth]{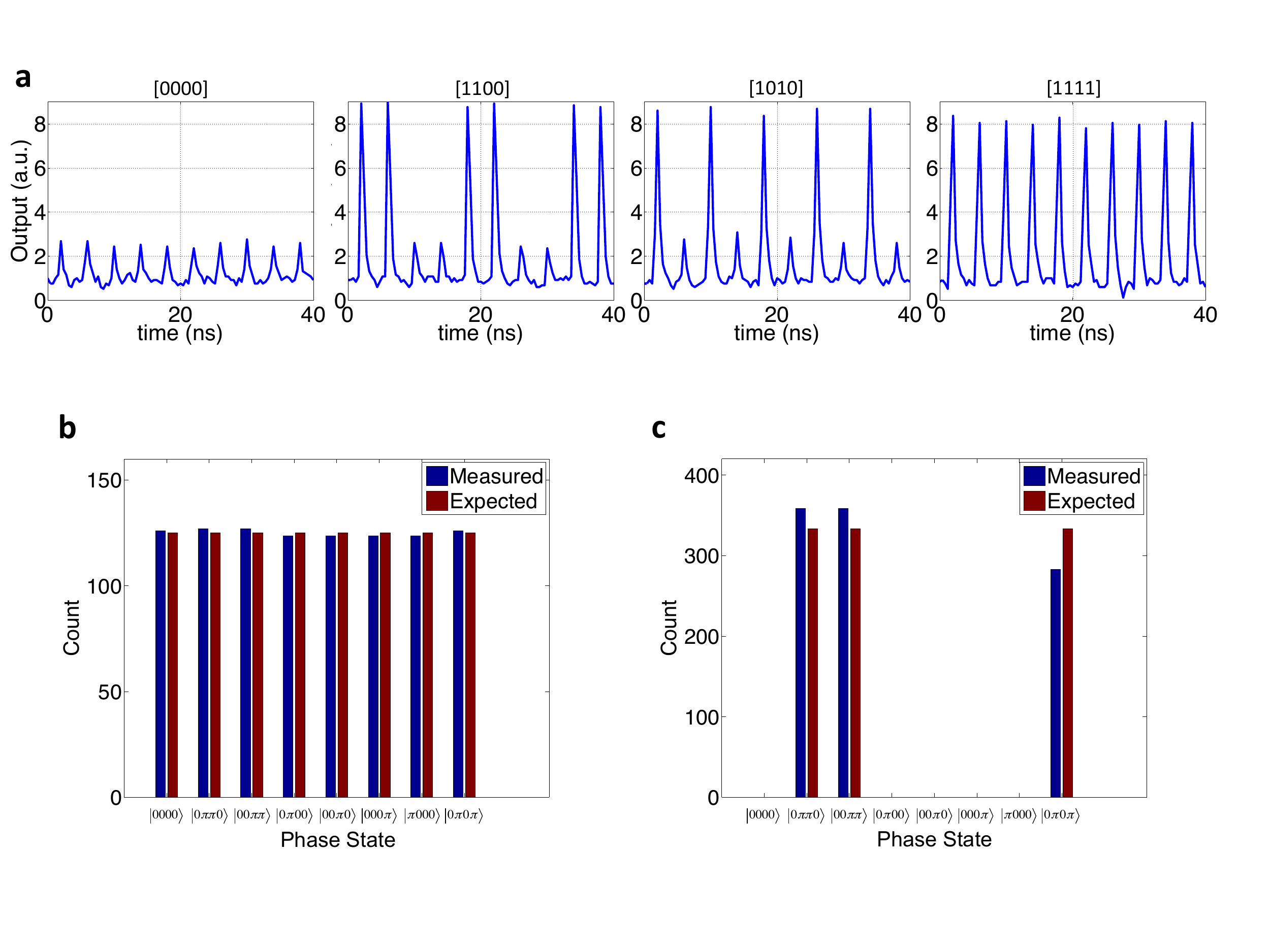}
\caption{ {\bf Fast detector results.} {\bf a}, Samples of the pulse patterns at the output of the interferometer. Each time the system is pumped above threshold, depending on the network configuration, one of these patterns is detected with a fast detector at the output of the interferometer.  {\bf b}, The histogram of the phase states when all the couplings are blocked, i.e. OPOs are independent, for 1000 runs. {\bf c}, The histogram of the phase states when all the delays are on and their phases are locked to $\pi$ for 1000 runs. The OPO network represents the MAX-CUT problem for a 4-vertex graph, and it only oscillates in the phase states corresponding to the answers of this problem. In {\bf b} and {\bf c} each phase state on the horizontal axis represents two complementary states, for example $|0000\rangle$ and $|\pi\pi\pi\pi\rangle$, because the pulse trains of the interferometer output are the same for such phase states. Also, in the measurements, no time reference is used, and only the pattern of the pulses are used to infer the phase states; the measured entries for $|0\pi00\rangle$, $|00\pi0\rangle$, $|000\pi\rangle$, and $|\pi000\rangle$ are the same and equal to the number of measured [1100] patterns divided by 4. Likewise, the entries for $|0\pi\pi0\rangle$ and $|00\pi\pi\rangle$ are the number of [1010] patterns divided by 2. }
\label{fig:fast_det}
\end{figure} 

When a fast detector is used at the interferometer output and all the couplings are blocked, after the 4-OPO system is turned on, one of the four possible pulse patterns (Supplementary Information) is detected as shown in Fig. \ref{fig:fast_det}a. In these patterns, the pulses are separated by 4 ns --the repetition period of the pump-- and each pulse has either a low or a high intensity corresponding to destructive or constructive interference of the consecutive OPOs, respectively. The low-level pulses are non-zero because of the diffraction mismatch of the interferometer arms. For the case of no coupling, the 4-OPO system is restarted 1000 times, and the histogram of the eight phase states is shown in Fig. \ref{fig:fast_det}b. For each restart, the pulse train is detected and the corresponding phase state is inferred. The result indicates uniform distribution of the phase states confirming that four temporally separated OPOs are operating independently in the same resonator.

To verify that the OPO network is capable of solving the NP-hard MAX-CUT problem, the histogram of the OPO states is measured for the case of all the delays locked to the phase of $\pi$ for 1000 trials. The result is depicted in Fig. \ref{fig:fast_det}c. No excited phase state is detected, and the distribution of the answer states is close to uniform. The error rate of computation is less than $10^{-3}$ which is limited by the length of measurements.

In conclusion, we demonstrate a coherent network of time-division-multiplexed degenerate OPOs that is capable of solving the NP-hard MAX-CUT problem for $N$=4. The practical scalability enabled by multi-pulse operation in a long ring cavity, simple mapping of the OPO bi-stable phase states to the Ising spin, and intriguing quantum phase transition at the threshold can make the OPO network suitable for obtaining approximate solutions for large-scale NP-hard Ising problems with reasonable accuracy and speed. Time division multiplexing enables increasing the size of the network by increasing the cavity round-trip time ($T_{cavity}=NT_R$), and implementing $N-1$ delay lines, with each delay corresponding to one coupling at a given time slot (see Supplementary Information for a practical implementation). To program a large-scale machine to arbitrary Ising problems, intensity modulators are needed to turn on and off each delay as well as phase modulators to choose either in-phase or out-of-phase couplings. Exploiting optical fiber technologies and planar light wave circuits can enable implementation of compact large Ising machines.

\section*{Acknowledgements}
The authors would like to thank S. E. Harris, H. Mabuchi, M. Armen, S. Utsunomiya, S. Tamate, K. Yan, and Y. Haribara, for their useful discussions, and K. Ingold, C. W. Rudy, C. Langrock, and K. Urbanek for their experimental supports. The work is supported by the FIRST Quantum Information Processing project.

\section*{Contributions}
A.M. and Y.Y. conceived the idea and designed the experiment. A.M. and K.T. carried out the experiment. Z.W. performed the numerical simulations. Y. Y. and R.L.B. guided the work. A.M. wrote the manuscript with input from all authors.

\newpage
\section*{Supplementary Information}
\setcounter{page}{1}
\renewcommand{\thepage}{S-\arabic{page}}

\setcounter{equation}{0}
\renewcommand{\theequation}{S-\arabic{equation}} 

\setcounter{figure}{0}

\section{Principle of Operation}

\begin{figure}[htbp]
\centering
\makeatletter 
\renewcommand{\thefigure}{S\@arabic\c@figure} 
\includegraphics[width=.7\textwidth]{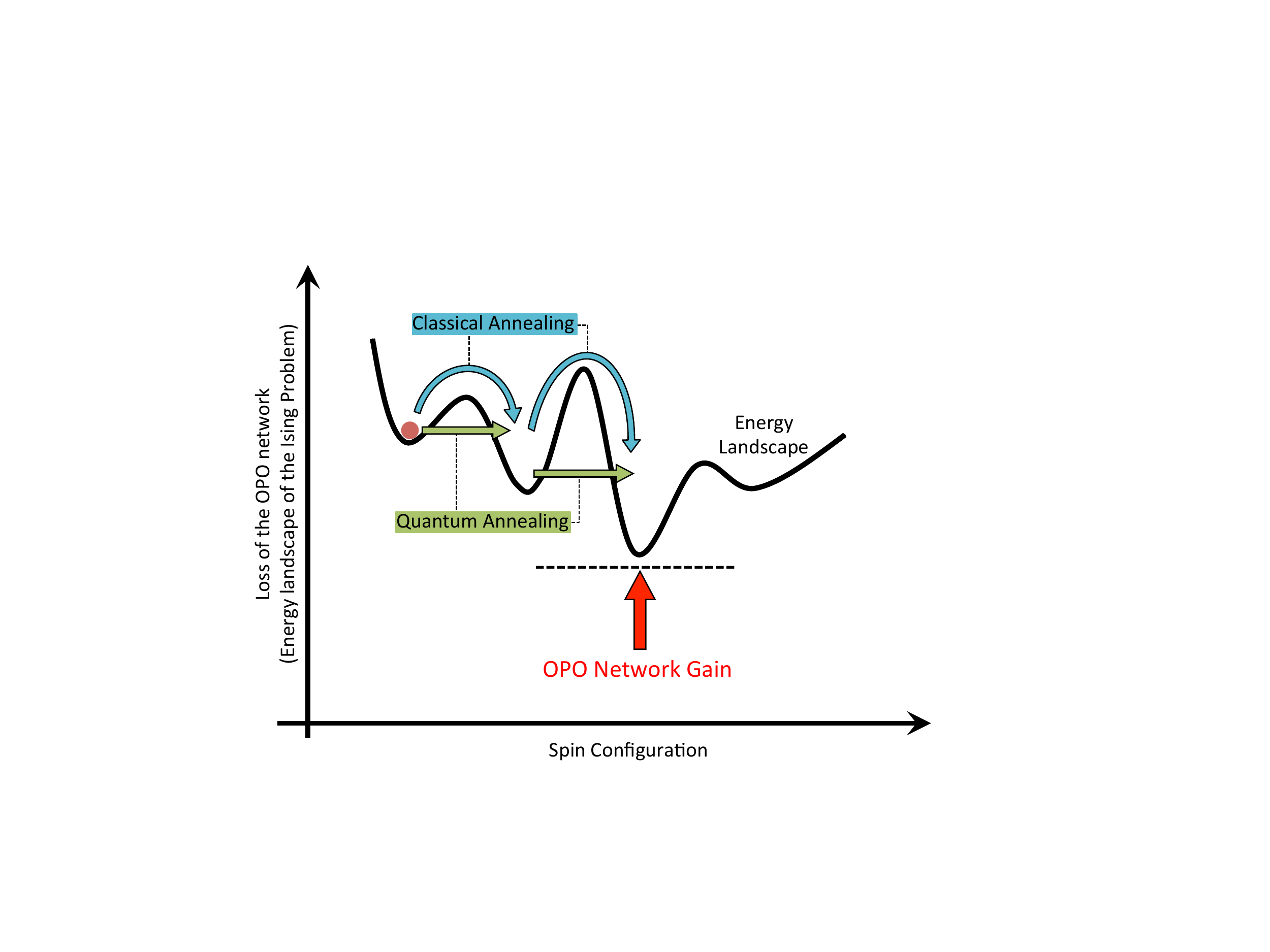}
\caption{ {\bf Comparison of the search mechanism between the OPO network, classical simulated anneaing, and quantum annealing.}  }
\label{fig:princ}
\end{figure}

The proposed computational concept is in sharp contrast with the classical and quantum annealing techniques. The cartoon of Fig. \ref{fig:princ} illustrates the challenge in classical and quantum annealing to go from a metastable excited state to the ground state through multiple thermal hopping or quantum mechanical tunneling through many metastable states. Classical simulated annealing employs a downward vertical search, in which the temperature is repeatedly decreased and increased until the ground state is found. Quantum annealing exerts a horizontal search in the energy landscape with quantum tunneling. Therefore, with these methods, the computational time of finding the ground state increases with the increase in the number of metastable excited states or local minima. In contrast, the OPO-based Ising machine searches for the ground state in an upward direction. The total energy, i.e. the ordinate in Fig. \ref{fig:princ}, is now replaced by the network loss. The ground state (optimum solution) has a minimum loss as shown in the figure. If we put a parametric gain (G) into such a network and increase it gradually, the first touch to the network loss happens at the ground state, which results in the single-mode oscillation of the ground state spin configuration. At the pump rate above this threshold point, the parametric gain is clamped at the same value of $G=L_{min}$ due to nonlinear gain saturation, so that all the other modes including local minima stay under the oscillation threshold. If we use the terminology of ``negative temperature" to represent the parametric gain, the mentioned upward search corresponds to the heating process from $T=-\infty$ for zero gain toward $T=-0$ for high gain. In this sense, the OPO machine is a ``heating machine" while the classical simulated annealing is a ``cooling machine." 

\section{Theoretical Modeling and Numerical Test of OPO Network}
To simulate a network of $N$ degenerate OPOs, we can start with the quantum mechanical Fokker-Planck equation (Q-FPE) for a single OPO using the generalized P-representation \cite{qpfe}:
\begin{align}
\label{eq:qfpe}
\frac{d}{dt}P(a_s,b_s)=\left\{ \frac{\partial}{\partial a_s}[a_s-(p-a_s^2)b_s)] + \frac{\partial}{\partial b_s}[b_s-(p-b_s^2)a_s)]  \right\} P(a_s,b_s) \notag\\
+\frac{1}{2A_s^2} \left[ \frac{\partial^2}{\partial a_s^2} (p-a_s^2)+\frac{\partial^2}{\partial b_s^2} (p-b_s^2) \right] P(a_s,b_s).
\end{align}
Here $a_s=\alpha_s/A_s$ and $b_s=\beta_s/A_s$ are the normalized eigenvalues for the off-diagonal coherent-state expansion, $|\alpha_s \rangle\langle\beta_s|$, of the density matrix,  $\gamma_s$ and $\gamma_p$ are the signal and pump photon decay rates, $A_s=(\gamma_p\gamma_s/2\kappa^2)^{1/2}$ is the oscillation field amplitude at a normalized pump rate $p=F_p/F_{th}=2$, $t=\frac{\gamma_s}{2}\tau$ is the normalized time, $F_p$ is the pump field amplitude, $F_{th}=\gamma_s\gamma_p/4\kappa$ is the threshold pump amplitude. The average amplitudes of in-phase and quadrature-phase components of the signal wave are obtained by
\begin{align}
\langle A_{s1} \rangle = A_s \langle a_s+b_s \rangle /2, \notag \\
\langle A_{s2} \rangle = A_s \langle a_s-b_s\rangle /2i.
\end{align}

Equation \ref{eq:qfpe} can be cast into the c-number Langevin equation (C-LGE) for the in-phase and quadrature-phase components of the signal field via the Kramers-Moyal expansion \cite{km}:
\begin{align}
\label{eq:csde}
dc=(-1+p-c^2-s^2)c \; dt + \frac{1}{A_s}\sqrt{c^2+s^2+\frac{1}{2}} \;dW_1,\notag\\
ds=(-1-p-c^2-s^2)s \; dt + \frac{1}{A_s}\sqrt{c^2+s^2+\frac{1}{2}} \;dW_2,
\end{align}
where $c=\frac{1}{2}(a_s+b_s)$ and $s=\frac{1}{2i}(a_s-b_s)$, and $dW_1$ and $dW_2$ are two independent Gaussian noise processes which represent the incident vacuum fluctuations at signal frequency $\omega_s$ and pump frequency $\omega_p$.

The equivalence of the Q-FPE (\ref{eq:qfpe}) and the C-LGE (\ref{eq:csde}) can be confirmed by comparing the squeezing and anti-squeezing characteristics of the two quadrature components using
\begin{align}
\langle \Delta A_{s1}^2\rangle = A_s^2 [ \langle (a_s+b_s)^2\rangle +1]/4 - \langle A_{s1}\rangle^2, \notag\\
\langle \Delta A_{s2}^2\rangle = -A_s^2 [ \langle (a_s-b_s)^2\rangle -1]/4 - \langle A_{s2}\rangle^2, 
\end{align}
for the Q-FPE and
\begin{align}
\langle \Delta A_{s1}^2\rangle = A_s^2 (\langle c^2 \rangle - \langle c \rangle ^2), \notag\\
\langle \Delta A_{s2}^2\rangle = A_s^2 (\langle s^2 \rangle - \langle s \rangle ^2),  
\end{align}
for the C-LGE. As shown in Fig. 1 in [7], the degrees of squeezing and anti-squeezing obtained by the two methods completely agree at a pump rate across the threshold ($p=1$).

The Q-FPE and C-LGE for an injection-locked laser oscillator \cite{laser_inj} can be extended to the mutually coupled degenerate OPOs using equation \ref{eq:qfpe} or \ref{eq:csde}. The resulting C-LGEs for a network of degenerate OPOs are given by
\begin{align}
\frac{d}{dt}c_j = \left \{ [-1+p-(c_j^2+s_j^2)]c_j + \sum\limits_{l=1}^N  \xi_{jl} c_l \right \} dt \notag\\
+\frac{1}{A_s}\sqrt{c_j^2+s_j^2+1/2} \; dW_{j1}, \notag\\
\frac{d}{dt}s_j = \left \{ [-1-p-(c_j^2+s_j^2)]s_j + \sum\limits_{l=1}^N  \xi_{jl} s_l \right \} dt \notag\\
+\frac{1}{A_s}\sqrt{c_j^2+s_j^2+1/2} \; dW_{j2}.
\end{align}
Here $c_j$ and $s_j$ are the normalized amplitudes of two quadrature components of the $j$-th OPO which corresponds to $c$ and $s$ in Eq. \ref{eq:csde}.

Performance of the proposed network of degenerate OPOs as an Ising machine is tested against the NP-hard MAX-CUT problems on cubic graphs for $N=4$ to $N=20$ and on random graphs for $N=800$ to $N=20000$. For a graph with $N$ vertices, the $2N$ C-LGEs are solved by the Dormand-Prince method as the differential equation solver \cite{dormand-prince}, in which adaptive integration strength is introduced by evaluating the local truncation error. 

\begin{figure}[htbp]
\centering
\makeatletter 
\renewcommand{\thefigure}{S\@arabic\c@figure} 
\includegraphics[width=.8\textwidth]{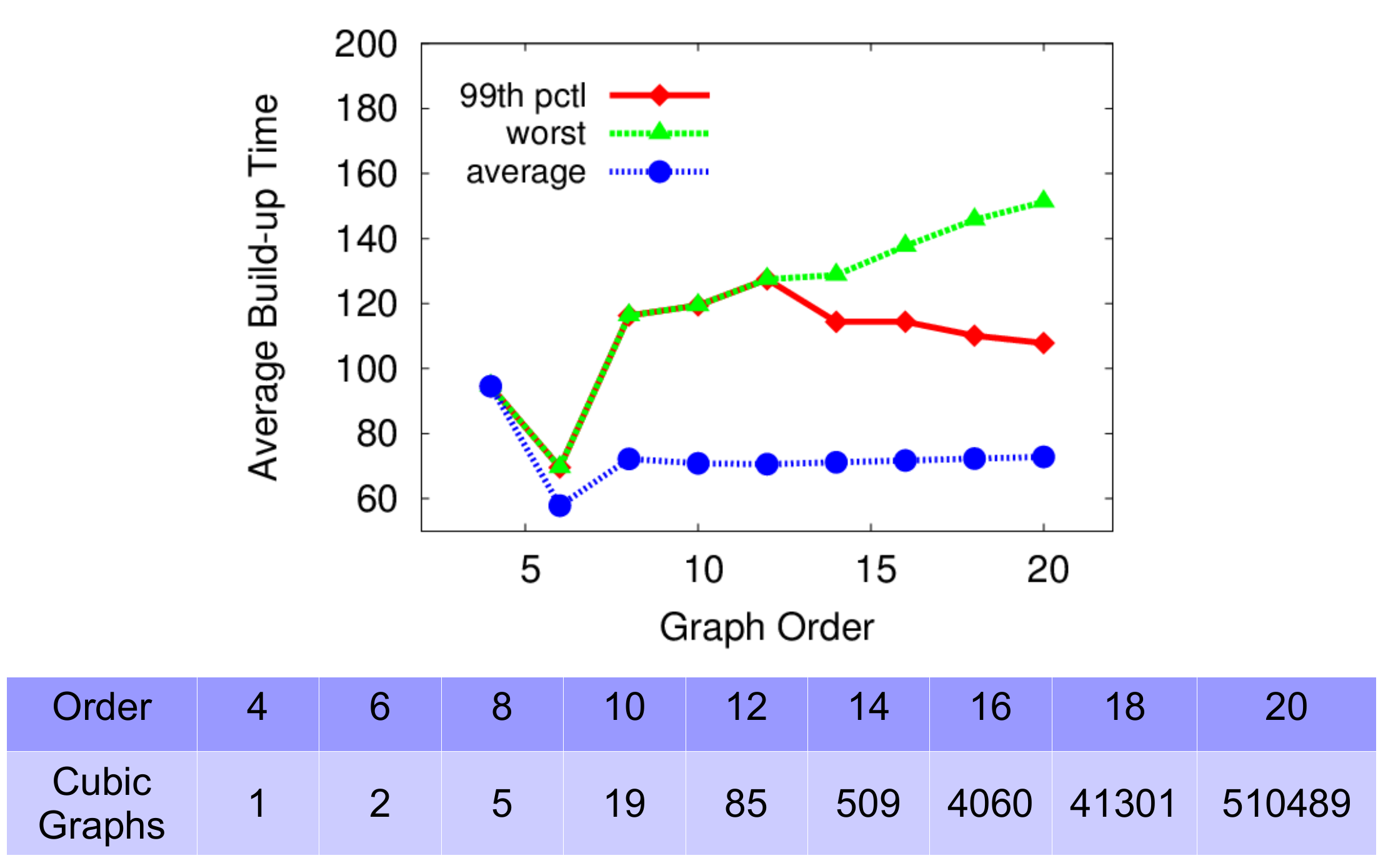}
\caption{ {\bf Numerical results of the build-up time of the OPO networks.} $p = 1.1$ and $\xi = -0.1$. The table indicates the number of cubic graphs for different orders.  }
\label{fig:buildup_time}
\end{figure}

Figure \ref{fig:buildup_time} shows the normalized build-up time $t=\frac{\gamma_s}{2}\tau$ when the OPO network reaches the steady state oscillation conditions after an above-threshold pump rate ($p=1.1$) is turned on versus the graph order $N$. We have numerically tested all graphs, for instance a total number of 510489 graphs are studied for $N$=20. Most of the build-up time (up to 99\% of all graphs) is independent of the graph order $N$ and is on the order of $t\simeq$100, as shown in Fig. \ref{fig:buildup_time}. Only slight increase is observed for the worst case as shown in green triangles. Therefore, an actual computational time is determined mainly by the success probability for obtaining a ground state. Since the proposed OPO network is a stochastic machine driven by quantum noise, the success probability is always smaller than one.

\begin{table}
\makeatletter
\renewcommand{\thetable}{S\@arabic\c@table} 
\caption{\bf{Summary of the OPO network in solving the MAX-CUT problems on cubic graphs.} }
\centering
\begin{tabular}{ccccccccc}
\hline\hline 
 & \multicolumn{8}{c}{Order} \\ & 4 & 6 & 8 & 10 & 12 & 14 & 16 & 20 \\
\hline
$q_{min}$ & 0.93 & 1.00 & 0.41 &  0.54 & 0.52 & 0.37 & 0.33 & 0.11 \\
$p_{opt}$ & 1.05 & 1.30 & 1.30 & 1.30 & 1.00 & 0.85 & 0.85 & 0.82 \\
$q_{opt}$ & 1.00 & 1.00 & 0.70 & 0.74 & 1.00 & 1.00 & 1.00 & 0.74 \\
 
\hline\hline
\end{tabular}
\label{tbl:cubic_graphs}
\end{table}

Table \ref{tbl:cubic_graphs} summarizes the performance of the OPO network in solving MAX-CUT problems on cubic graphs. Here, $q_{min}$ denotes the worst-case success probabilities at a fixed pump rate $p = 1.1$ and coupling coefficient $\xi = -0.1$, $p_{opt}$ denotes the optimal pump rate for each worst-case instance, at which the optimum success probability $q_{opt}$ is achieved under the same coupling coefficient $\xi = -0.1$. The success probability at the optimum pump rate for the worst instance is independent of the graph order and ranges from $0.7\sim1.0$.

\begin{table}
\makeatletter
\renewcommand{\thetable}{S\@arabic\c@table} 
\caption{ {\bf Performance of the OPO network in solving the MAX-CUT problems on sample G-set graphs.} $V$ is the number of vertices in the graph, $E$ is the number of edges, $U_{SDP}$ is the optimal solution to the semidefinite relaxation of the MAX-CUT problem, and $T$ is the average computation time of the OPO network normalized to the cavity photon lifetime. To make comparisons with the Goemans-Williamson algorithm, every cut value $O$ generated from the network is normalized according to $(O + E_{neg})/(U_{SDP}+E_{neg})$, where $E_{neg} \geq 0$ is the number of negative edges. $O_{max}$ and $O_{avg}$ are the best and the average values in 100 runs, respectively.}
\centering
\begin{tabular}{cccccccc}

\hline\hline 
Graph & $V$ & $E$ & $U_{SDP}$ & $O_{max}$ & $O_{avg}$ & $T$ \\    
\hline
G1	&	800	&	19176	&	12083	&	0.9591	&	0.9516	&	498.82	\\
G6	&	800	&	19176	&	2656	&	0.9559	&	0.9506	&	471.06	\\
G11	&	800	&	1600	&	629	&	0.9384	&	0.9254	&	406.24	\\
G14	&	800	&	4694	&	3191	&	0.9367	&	0.9274	&	498.26	\\
G18	&	800	&	4694	&	1166	&	0.9308	&	0.9223	&	430.24	\\
G22	&	2000	&	19990	&	14136	&	0.9349	&	0.9277	&	768.34	\\
G27	&	2000	&	19990	&	4141	&	0.9321	&	0.9270	&	780.18	\\
G32	&	2000	&	4000	&	1567	&	0.9328	&	0.9260	&	467.42	\\
G35	&	2000	&	11778	&	8014	&	0.9264	&	0.9202	&	602.34	\\
G39	&	2000	&	11778	&	2877	&	0.9214	&	0.9152	&	539.9	\\
G43	&	1000	&	9990	&	7032	&	0.9373	&	0.9309	&	542.92	\\
G48	&	3000	&	6000	&	6000	&	0.9463	&	0.9292	&	762.34	\\
G51	&	1000	&	5909	&	4006	&	0.9333	&	0.9242	&	491.68	\\
G55	&	5000	&	12498	&	11039	&	0.9070	&	0.9009	&	903.46	\\
G57	&	5000	&	10000	&	3885	&	0.9305	&	0.9259	&	648.72	\\
G59	&	5000	&	29570	&	7312	&	0.9114	&	0.9074	&	583.54	\\
G60	&	7000	&	17148	&	15222	&	0.9037	&	0.8995	&	918.98	\\
G62	&	7000	&	14000	&	5431	&	0.9295	&	0.9256	&	719.74	\\
G64	&	7000	&	41459	&	10466	&	0.9129	&	0.9092	&	666.16	\\
G65	&	8000	&	16000	&	6206	&	0.9284	&	0.9252	&	757.28	\\
G66	&	9000	&	18000	&	7077	&	0.9285	&	0.9251	&	786.9	\\
G67	&	10000	&	20000	&	7744	&	0.9285	&	0.9260	&	732.14	\\
G70	&	10000	&	9999	&	9863	&	0.9433	&	0.9379	&	687.28	\\
G72	&	10000	&	20000	&	7809	&	0.9284	&	0.9256	&	748.3	\\
G77	&	14000	&	28000	&	11046	&	0.9281	&	0.9256	&	842.24	\\
G81	&	20000	&	40000	&	15656	&	0.9268	&	0.9250	&	980.54	\\
\hline\hline
\end{tabular}
\label{tbl:G_graphs}
\end{table}

The performance of the OPO network in solving the MAX-CUT problems has also been examined on 71 benchmark instances of the so-called G-set graphs when $p = 1.1$ and $\xi = -0.1$. These instances are randomly constructed by a machine-independent graph generator written by G. Rinaldi, with the number of vertices ranging from 800 to 20000, edge density from 0.02\% to 6\%, and geometry from random, almost planar to toroidal. The outcomes of running the OPO network 100 times for sample G-set graphs are summarized in Table. \ref{tbl:G_graphs}. It can be easily seen that both the best and average outputs of the OPO network are about 2 - 6\% better than the 0.878-performance guarantee of the celebrated Goemans-Williamson algorithm based on semidefinite programming (SDP) \cite{GW95}. Since the differences between the best and the average values are within 1\% for most of the instances, reasonable performance is expected for the OPO network even in a single run, which makes the OPO network favorable for applications when response time is the utmost priority. In addition, there is further room to improve the performance, for example by applying local improvement to the raw outcomes of the OPO network and operating the OPO network under optimum pump rate of $p$ and coupling strength of $\xi$.

\begin{figure}[htbp]
\centering
\makeatletter 
\renewcommand{\thefigure}{S\@arabic\c@figure} 
\includegraphics[width=.7\textwidth]{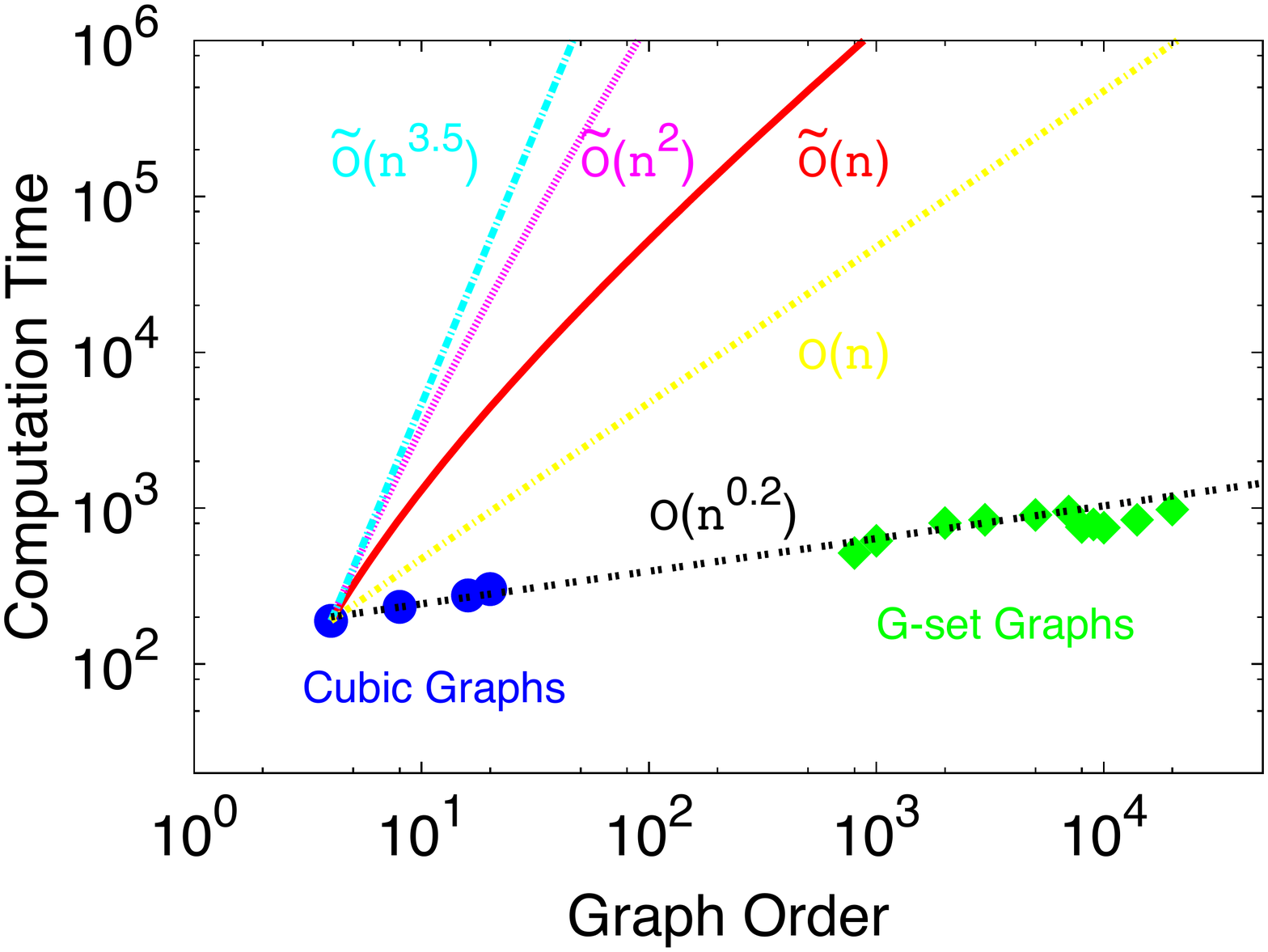}
\caption{ \bf{Average computational time of the OPO network in solving MAX-CUT problems on the worst-case cubic and G-set graphs.} $p = 1.1$ and $\xi = -0.1$. }
\label{fig:time_scaling}
\end{figure}

The average computational time of the OPO network in solving the MAX-CUT problem on the worst-case cubic and G-set graphs is displayed in Fig. \ref{fig:time_scaling}. The growth of the computation time fits well to a sub-linear function $O(N^{0.2})$. Computational complexities of best-known algorithms for solving the SDP in the Goemans-Williamson algorithm are also plotted. If a graph with $N$ vertices and $m$ nodes is regular, the SDP can be approximately solved in almost linear time as $\tilde{O}(m) = O(m\log^2(N)\epsilon^{-4})$ using the matrix multiplicative weights method \cite{AK07}, where $\epsilon$ represents the accuracy of the obtained solution. This behavior is shown by the red solid line in Fig. \ref{fig:time_scaling}. However, slower algorithms are required for general graphs. If the edge weights of the graph are all non-negative, the fastest algorithm runs in $\tilde{O}(Nm) = O(Nm\log^2(N)\epsilon^{-3})$ time based on a Lagrangian relaxation-based method \cite{KL96}. This computational time is plotted by the pink solid line. For graphs with both positive and negative edge weights, the SDP is commonly solved using the interior-point method which scales as $\tilde{O}(N^{3.5}) = O(N^{3.5}\log(1/\epsilon))$ \cite{Ali95}. This general computational time for SDP is shown by the blue dashed line. Since the OPO network is applicable to all types of graphs, the sub-linear scaling of the computation time gives it a huge advantage over the SDP algorithm in solving large-scale instances. For instance, the OPO network outputs a solution for the graph G81 with 20000 vertices and 40000 edges in a normalized time of $t\simeq$980, which corresponds to the actual time of $\tau=2\gamma_s^{-1}t \simeq 1\times 10^{-2}$ sec, since the photon lifetime in the large-scale fiber-based OPO network , which can handle this size of problem, is $\sim 6 \times 10^{-6}$ sec (see the last section in this Supplementary Information). This computational time is compared to the time required to run the SDP (interior-point method) using a 1.7 GHz core i7 machine of about $\sim1\times10^5$ sec, which is seven orders of magnitude larger than $\sim1\times 10^{-2}$ sec for the OPO network.

\section{Measurement of Phase States}
For the 4-OPO system, the one-bit delay interferometer is used for measuring the phase states. Table \ref{table:phase_states} shows the phase states and corresponding output pulse trains in the first and second columns. Since the complementary phase states result in the same output, for the 16 possible phase states 8 different pulse trains can occur. In the measurements with a fast detector, no time reference is used, therefore only four different pulse patterns can be detected. If a slow detector is used, the output will be proportional to the number of ones in the pulse train. Therefore, there will be three distinct output average intensity levels namely $I_m$, $I_m/2$, and $0$, as shown in the third column of the table.

\begin{table}
\makeatletter
\renewcommand{\thetable}{S\@arabic\c@table} 
\caption{{\bf Measurement of phase states.} All 16 possible phase states of the 4-OPO system, the corresponding pulse trains at the output of the unequal-arm interferometer, and the detected intensity using a slow detector. Highlighted rows are the answers of MAX-CUT.}
\label{table:phase_states}
\begin{center}
\begin{tabular}{ c  c  c  c  c }
  \hline  \hline
    \rowcolor{white}                     
  \bf{} & \bf{Interferometer} &  \bf{} \\
    \rowcolor{white}
    \bf{State} & \bf{Pulse Train} &  \bf{Slow Detector} \\
\hline
  $|0000\rangle$ & [1111] &  $I_m$ \\ 
  $|\pi000\rangle$ & [0110] &  $I_m/2$ \\ 
  $|0\pi00\rangle$ & [0011] & $I_m/2$ \\ 
   \rowcolor{yellow}
  $|\pi \pi00\rangle$ & [1010] & $I_m/2$ \\ 
  $|00\pi0\rangle$ & [1001] & $I_m/2$ \\ 
   \rowcolor{yellow}
  $|\pi0\pi0\rangle$ & [0000] & $0$ \\ 
  \rowcolor{yellow}
  $|0\pi\pi0\rangle$ & [0101] & $I_m/2$ \\ 
  $|\pi\pi\pi0\rangle$ & [1100] & $I_m/2$ \\ 
  $|000\pi\rangle$ & [1100] & $I_m/2$ \\ 
  \rowcolor{yellow}
  $|\pi00\pi\rangle$ & [0101] & $I_m/2$ \\ 
  \rowcolor{yellow}
  $|0\pi0\pi\rangle$ & [0000] & $0$ \\ 
  $|\pi\pi0\pi\rangle$ & [1001] & $I_m/2$ \\ 
  \rowcolor{yellow}
  $|00\pi\pi\rangle$ & [1010] & $I_m/2$ \\ 
  $|\pi0\pi\pi\rangle$ & [0011] & $I_m/2$ \\ 
  $|0\pi\pi\pi\rangle$ & [0110] & $I_m/2$ \\ 
  $|\pi\pi\pi\pi\rangle$ & [1111] &  $I_m$ \\ \hline \hline

\end{tabular}
\end{center}
\end{table}

\section{Details of Experimental Setup}
The OPO design shares similarities with the experiment reported in \cite{rudy}. The ring resonator of the OPO illustrated in Fig. 1b, has a round trip time of 16 ns (a perimeter of $\sim$4.8 m). The setup has two more flat mirrors than the schematic (corresponding to a folded bow tie configuration). All the flat mirrors, except M$_1$ are gold coated with enhancement dielectric coatings at 2 $\mu$m. One of the flat gold mirrors is placed on a translation stage with piezoelectric actuator (PZT). The dielectric mirror (M$_1$) has a coating which is antireflective at the pump wavelength with less than 0.2\% reflection, and is highly reflective ($\sim$99\%) from 1.8 $\mu$m to 2.4 $\mu$m. The curved mirrors (M$_3$ and M$_4$) have 50-mm radius of curvature and are unprotected gold coated mirrors. The angle of incidence on these mirrors is 4$^\circ$, which is chosen to  compensate the astigmatism introduced by the Brewster-cut nonlinear crystal, and results in $\sim$1 mm of cavity stability range for the spacing between the curved mirrors. The signal beam has a waist radius of 8.3 $\mu$m (1/e$^2$ intensity) at the center of the crystal. 

The 1-mm long, Brewster-cut, MgO:PPLN crystal has a poling period of 31.254 $\mu$m, which is designed to provide degenerate parametric gain for a pump at 1035 nm with type 0 phase matching (e$\rightarrow$e+e) at 373 K temperature. The crystal operates at room temperature in the OPO, and even though the phase matching condition is not optimal for the pump (centered at 1045 nm), degenerate operation is achieved by length tuning of the cavity. 

Input and output coupling of the signal are achieved with 2-$\mu$m thick nitrocellulose pellicles to avoid dispersion in the cavity and etalon effects. In Fig. 1b, the three pairs of ``OC" and ``IC" are uncoated pellicles (with $2-6$\% power reflection ), the ``OC" for the main output is coated (with $\sim$15\% power reflection at 2090 nm). The beam splitter in the interferometer (BS) is the same coated pellicle. For stabilizing the OPO cavity, another uncoated pellicle is used as an output coupler in the resonator (not shown in the schematic).

The pump is a free-running mode-locked Yb-doped fiber laser (Menlo Systems Orange) producing $\sim$80 fs pulses centered at 1045 nm with a repetition rate of 250 MHz, and maximum average power of $>$ 1 W. The filter is a long pass filter at 1850 nm on a Ge substrate to eliminate the pump and transmit the signal. 

Gradual pumping is achieved by the chopper, as it causes a rise time (10-90\% power) of 180 $\mu$s for introducing the pump. The cavity photon lifetime for the signal is estimated to be 60 ns, and the network is pumped $\sim$2.2 times above threshold.

\subsection{Servo Loops}
Five feedback servo controllers are used to stabilize the length of the cavity, the phase of the delay lines, and the arm-length difference of the interferometer. All the controllers are based on ``dither-and-lock" scheme, where a slight modulation (less than 10 nm amplitude at a frequency between 5 and 20 kHz) is applied to a fast PZT, and the error signal is generated electronically by mixing the detector output and the modulated signal [13]. Identical electronic circuits are used with a controller 3-dB bandwidth of 10 Hz.

For the delay lines, the interference of the pump at the other port of the input couplers are used as the input of the controller, and the controller locks the length to achieve destructive interference on the detector, which results in constructive interference on the other port that enters the cavity. The arm length difference of the interferometer is also locked similarly. No phase stabilization is required for the path from the OPO to the interferometer since all the OPO pulses experience the same path and phase change.

\subsection{Free-Running Pump}
Here we show that the servo controllers used in the experiment suffice for implementation of the Ising machine and no stabilization on the pump is required. Slow changes (within the response time of the controller) in the carrier-envelop offset frequency ($f_{CEO}^p$) and repetition rate ($f_R$) of the pump do not affect the operation of the Ising machine. Smooth changes in $f_R$ of the pump is intrinsically transferred to the signal since signal pulses are generated from pump pulses. However, the effects of changes in $f_{CEO}^p$ on the Ising machine require taking into account the intrinsic phase locking of the degenerate OPO as well as the role of servo controllers.

The primary task of the servo controller of the OPO is to maximize the output power by matching the roundtrip phase in the resonator to the pulse to pulse phase slip of the pump ($\Delta \phi_p$). The pulse to pulse phase slip is related to $f_{CEO}^p$ by:
\begin{equation}
f_{CEO}^p=\frac{\Delta \phi_p}{2\pi} f_R.
\end{equation}
Assuming the carrier fields for the pump and the signal pulses are defined as: $exp(j\omega_p t+\phi_p)$, and $exp(j\omega_s t+\phi_s)$, respectively, the phase-sensitive gain dictates [13]:
\begin{equation}
\phi_p=2\phi_s+\pi/2.
\end{equation}
Therefore, if the carrier phase of the pump changes by $\Delta \phi_p$ from one pulse to next, for a single OPO ($T_{cavity}=T_R$), the pulse to pulse phase slip of the signal follows that by:
\begin{equation}
\Delta \phi_s=\Delta \phi_p/2.
\end{equation}
This means that the phase slip of the signal pulses is locked to the phase slip of the pump with a factor of half, which consequently means the $f_{CEO}$ of the pump and signal are locked [13]; the servo loop provides feedback to the cavity to follow this phase slip and maximize the output power. A similar behavior also happens for a doubly-resonant OPO operating away from degeneracy, with a different ratio between the $f_{CEO}$s of the pump and signal \cite{opo_fceo}.

For $N$ OPOs in the cavity (i.e. $T_{cavity}=NT_R$), when all OPOs are in the same phase state, the pulse to pulse phase slip in the pump transfers to the OPO to OPO phase slip by a factor of half. Changing phase state from one OPO to another simply means adding $\pi$ to the phase slip. When a delay line is locked to the top of the interference fringe of the pump pulses, the phase change provided by the delay line at the pump wavelength compensates the pulse to pulse phase slip of pump, i.e. $\phi_D(\omega_p)= \Delta \phi_p$. At the signal wavelength, because $\omega_s=\omega_p/2$ this phase change in the delay line is half ($\phi_D(\omega_s)=\phi_D(\omega_p)/2$), which means that the servo controller compensates the OPO to OPO phase slip resulting from the $f_{CEO}^p$. 

Locking a delay line to top of the fringe of pump pulses corresponds to having either $0$ or $\pi$ phase change for the signal. This is also true for the interferometer. In the experiments, for all the servo controllers, we were able to precisely tune the length from one fringe to the next. This gave us the ability to try different configurations and find the desired coupling phases.

\section{OPO Characterization}
\begin{figure}[htbp]
\centering
\makeatletter 
\renewcommand{\thefigure}{S\@arabic\c@figure} 
\includegraphics[width=.8\textwidth]{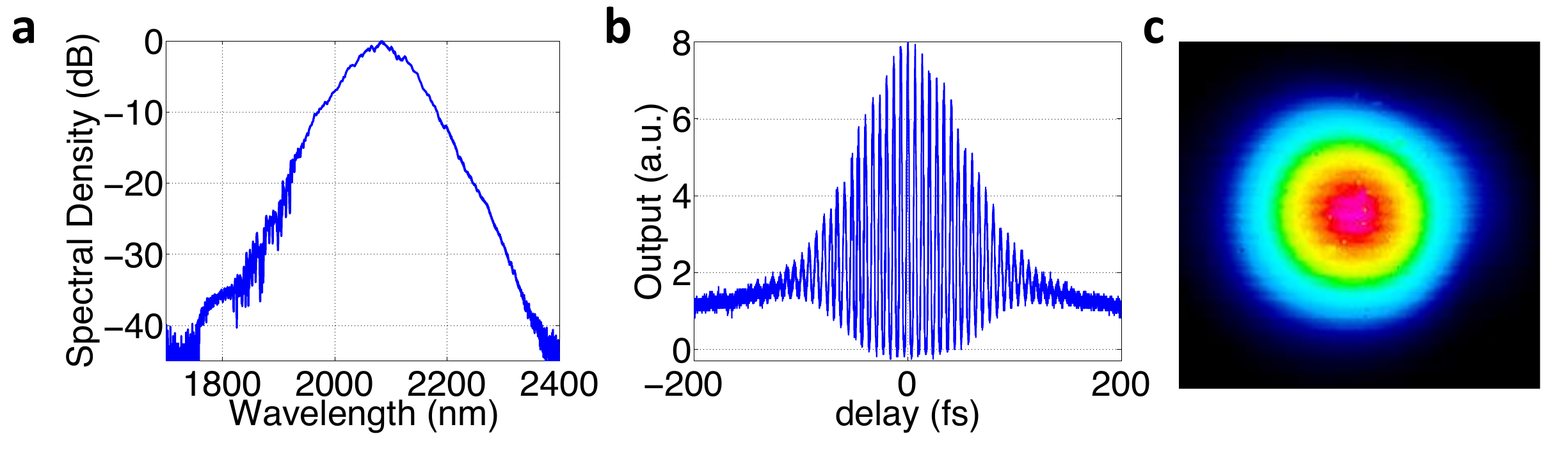}
\caption{ {\bf Summary of  the 4-OPO system operation.} {\bf a}, The output spectrum centered at 2090 nm with a 3-db bandwidth of 91 nm. {\bf b}, The interferometric autocorrelation trace of the output pulses with a FWHM of $\sim$120 fs suggesting Gaussian pulses of $\sim$85 fs. {\bf c}, The spatial beam profile of the OPO output at 2090 nm. }
\label{fig:OPO_char}
\end{figure}

When no out coupler is used in the cavity, the OPO has a threshold of 6 mW of pump average power. With all the output and input couplers in the cavity, the threshold reaches 135 mW. Oscillation at degeneracy and away from degeneracy can be achieved depending on the cavity length. The OPO is pumped with 290 mW and the main output of the OPO has 15 mW of average power at degeneracy centered at 2090 nm with the spectrum shown in Fig. \ref{fig:OPO_char}a. The interferometric autocorrelation of the signal pulses is shown in Fig. \ref{fig:OPO_char}b suggesting a pulse length of $\sim$85 fs. The spatial profile of the output beam is very close to Gaussian as shown in Fig. \ref{fig:OPO_char}c with a radius of $\sim$1 mm (at $1/e^2$ intensity). The average power of the signal in the delay lines are $\sim$2 mW, and the intracavity power is estimated to be $\sim$100 mW.

\section{Extended Slow-Detector Results}

\begin{figure}[htbp]
\centering
\makeatletter 
\renewcommand{\thefigure}{S\@arabic\c@figure} 
\includegraphics[width=\textwidth]{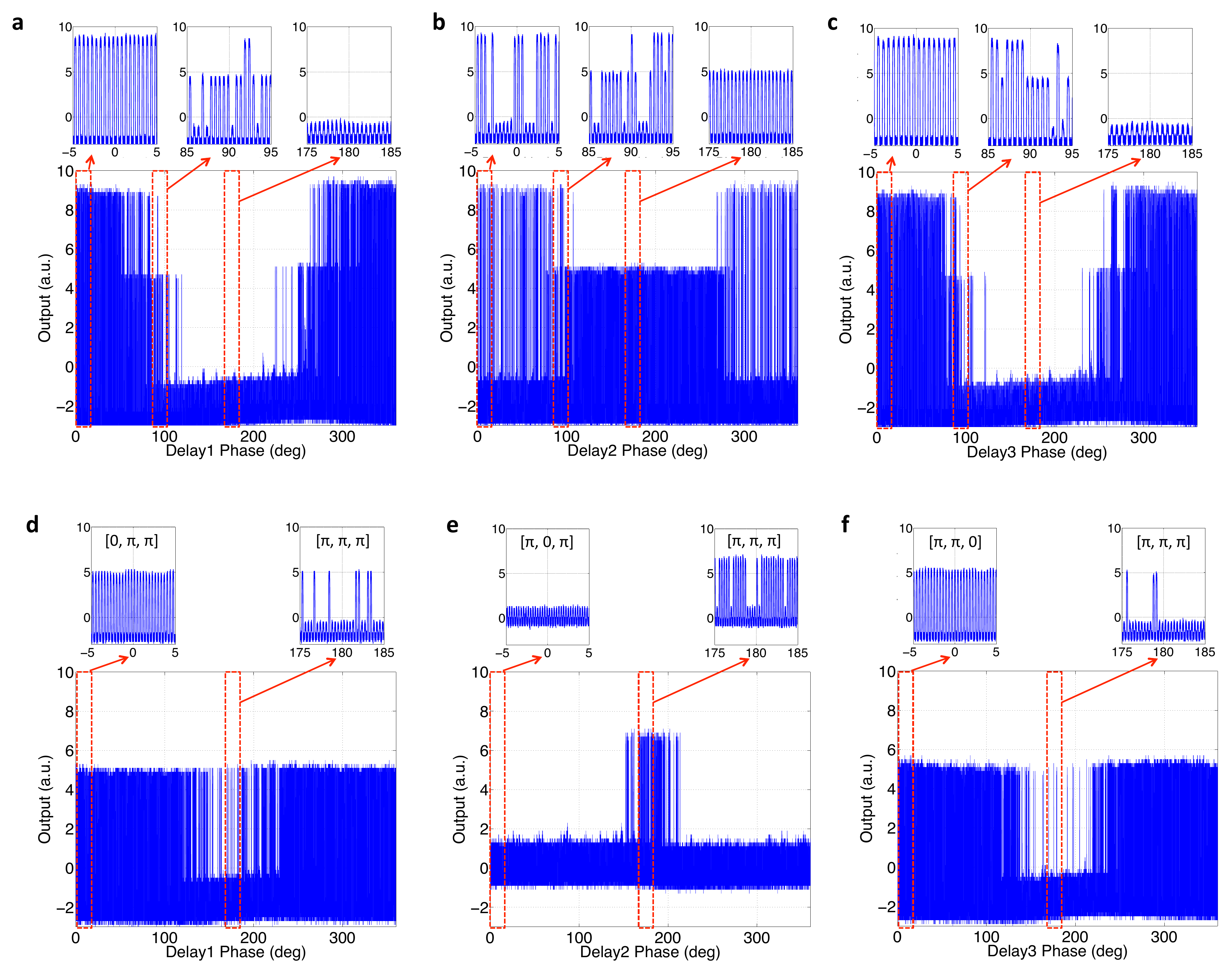}
\caption{ {\bf The output of the slow detector while the phase of one delay is scanned slowly with respect to the chopper, and the other delays are either blocked (a, b, c) or locked to $\pi$ (d, e, f).} {\bf a}, Delays 2 and 3 are blocked and phase of delay 1 is scanned. {\bf b}, Delay 1 and 3 are blocked and delay 2 is scanned. {\bf c}, Delays 1 and 2 are blocked and delay 3 is scanned. {\bf  d}, Phases of delays 2 and 3 are locked to $\pi$ and delay 1 is scanned; phase configuration of the network corresponds to $[0, \pi, \pi]$ at 0$^\circ$ and $[\pi, \pi, \pi]$ at 180$^\circ$. {\bf  e}, Phases of delays 1 and 3 are locked to $\pi$ and delay 2 is scanned; phase configuration of the network corresponds to $[\pi, 0, \pi]$ at 0$^\circ$ and $[\pi, \pi, \pi]$ at 180$^\circ$.  {\bf  f}, Phases of delays 2 and 3 are locked to $\pi$ and delay 1 is scanned; phase configuration of the network corresponds to $[\pi, \pi, 0]$ at 0$^\circ$ and $[\pi, \pi, \pi]$ at 180$^\circ$.  }
\label{fig:ex_slow_det}
\end{figure}

In Fig. \ref{fig:ex_slow_det} we show the results obtained using a slow detector for different combinations of couplings. Fig.  \ref{fig:ex_slow_det} a-c are obtained by scanning the phase of one delay line while the other delay lines are blocked. Delay 1 and 3 have similar effects, because they couple adjacent OPOs but in different directions (Fig. 2b). As shown in Fig.  \ref{fig:ex_slow_det} a and c, in-phase coupling by these delays results in the same phase state for all OPOs, and consequently high-intensity interferometer output ($I_m$); and out-of-phase coupling results in alternating phase states and consequently low-intensity interferometer output ($0$). 

In Fig. \ref{fig:ex_slow_det}b we show the interferometer output while the phase of delay 2 is scanned. When the coupling is in-phase, OPO 1 and 3 have the same phase state, and OPO 2 and 4 oscillate in the same phase state. However, these two pairs can either be the same or different, and therefore the output would be either $I_m$ or 0, as shown in Fig. \ref{fig:ex_slow_det}b around phase of zero. On the other hand, out-of-phase coupling of delay 2 results in constant output $I_m/2$ as shown in the same plot. Regenerative behavior of the OPO and its insensitivity to a wide range of phase change in the couplings are observed in these three plots.

When the network is configured to the MAX-CUT problem, we scanned the phase of the delays one by one, and the results are shown in Fig. \ref{fig:ex_slow_det}d-f. Different delay phase configurations and the expected outcomes are shown in Table \ref{table:phases}, where the last row corresponds to the MAX-CUT problem with all anti-ferromagnetic couplings, and for each of the other rows one of the delay phases is different. For each plot in  Fig. \ref{fig:ex_slow_det}d-f, the center of the plot, where the phase of the scanned delay is $\pi$, corresponds to the anti-ferromagnetic MAX-CUT problem. The outputs follow the expected outcome.

\begin{table}
\makeatletter 
\renewcommand{\thetable}{S\@arabic\c@table} 
\caption{Four configurations for the phases of the delays, the expected phase states of the 4-OPO system, and the expected outcome of the measurements; The last row is the phase configuration corresponding to the MAX-CUT problem, and in the rest of phase configurations one of the delays has a different phase. }
\label{table:phases}
\begin{center}
\begin{tabular}{| c | c | c | c | }
  \hline                        
  \bf{Phase of Couplings} & \bf{Expected Phase} &  \bf{} \\
    \bf{[D1, D2, D3]} & \bf{States} &  \bf{Slow Detector} \\
\hline
   $[\pi, 0, \pi]$ & $ |0\pi0 \pi \rangle, |\pi 0 \pi 0 \rangle$ &100\% in $0$ \\ \hline
  
    $[0, \pi, \pi]$ & $|00\pi \pi \rangle, |\pi00\pi \rangle,$  &100\% in $I_m/2$ \\ 
    & $|0\pi \pi 0 \rangle, |\pi \pi 00 \rangle$ &\\
    \hline 
  
  $[\pi, \pi, 0]$ & $|00\pi \pi \rangle, |\pi00\pi \rangle,$ &100\% in $I_m/2$ \\ 
 &  $|0\pi \pi 0 \rangle, |\pi \pi 00 \rangle$ & \\
  \hline
  
  \rowcolor{yellow}
  $[\pi, \pi, \pi]$ & $|00\pi \pi \rangle, |\pi00\pi \rangle, |0\pi \pi 0 \rangle, |\pi \pi 00 \rangle, $ &66.7\% in $I_m/2$, \\ 
  \rowcolor{yellow}
  & $|0\pi0 \pi \rangle, |\pi 0 \pi 0 \rangle$ & 33.3\% in $0$\\
  \hline
  
\end{tabular}
\end{center}
\end{table}

\section{Scalability}

Time division multiplexing facilitates scalability of the OPO network, and it benefits from intrinsically identical nodes in the network. It is also worth noting that for an Ising problem with $N$ sites, the number of possible couplings ($J_{ij}$) is $N^2-N$. However, in the time-division-multiplexed network only $N-1$ delay lines are required for realizing these couplings, which means that the physical size of the machine scales linearly with $N$. 

\begin{figure}[htbp]
\centering
\makeatletter 
\renewcommand{\thefigure}{S\@arabic\c@figure} 
\includegraphics[width=.7\textwidth]{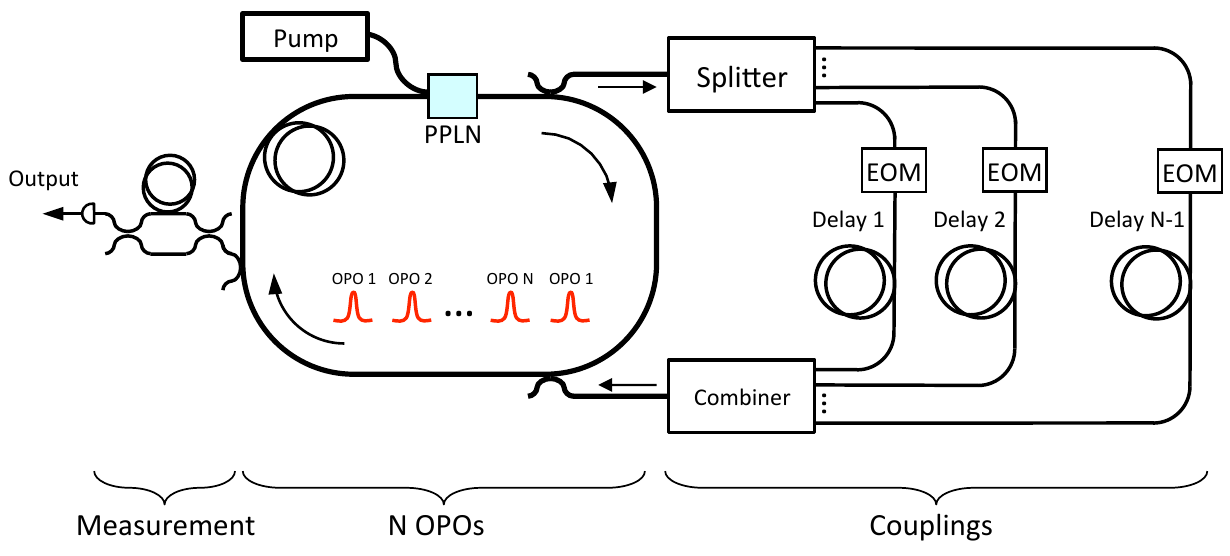}
\caption{ {\bf  Schematic of a fiber-based large-scale OPO network.} }
\label{fig:N_OPO_sch}
\end{figure}

A network of $N$ OPOs can be realized in a single ring resonator with a round-trip time of $T_{cavity}=NT_R$ ($T_R$ is the pulse-to-pulse interval), and constructing $N-1$ delay lines. Schematic of fiber-based implementation of such a machine is illustrated in Fig. \ref{fig:N_OPO_sch}. To avoid effects of nonlinearities and dispersion in optical fibers, picosecond pump pulses can be used in a long resonator and long delay lines comprising optical fiber components. As an example, for a pump with 10-GHz repetition frequency ($T_R=100$ ps), a resonator with 200 m of optical fiber results in 10000 temporally separated OPOs. An expected photon lifetime of such a fiber-based OPO network is about $\gamma_s^{-1} \simeq 6\times 10^{-6}$s, which promises a reasonably fast computational time for a MAX-CUT problem with $N=10000$. The main challenge is stabilizing the phases of all these fiber links. Development of extremely low-noise phase-stabilized long ($\sim$100-km) optical fibers \cite{fiber_locking} promises overcoming this challenge using the existing technologies. Moreover, the regenerative behavior of the degenerate OPO (as shown in Fig. 3b) suggests that the OPO-based Ising machine can tolerate relatively large phase noise in the couplings.

Similar to the 4-OPO network, each delay line provides a delay of equal to an integer multiple of the repetition period ($mT_R$), and is responsible for multiple of the Ising coupling terms in the form of $J_{(i)(i+m)}$. In one delay line, each of these couplings happens at one time slot, and one can use electrooptic phase and amplitude modulators (EOM) to synchronously switch the delay on and off depending on whether the corresponding coupling term is zero or not. This can be extended to synchronously controlling the phases and strengths of the couplings through the delay lines, and hence programming any arbitrary Ising problem on the machine.

\end{document}